\documentstyle[sprocl]{article}
\def\PRD{{\em Phys. Rev.} D}

\def\be{\begin{equation}}
\def\ee{\end{equation}}
\def\bea{\begin{eqnarray}}
\def\eea{\end{eqnarray}}

\begin{document}
\title{Non-Gaussian Spectra in the Cosmic Microwave Background}
\author{ Pedro G. Ferreira \footnote{In collaboration with Jo\~{a}o
Magueijo  (Imperial College)}}
\address{Center for Particle Astrophysics, 301 Leconte Hall,
\\ University of California, Berkeley CA 94720, USA}
%
\maketitle\abstracts{We propose a set of new statistics which can
be extracted out of the angular distribution of the Fourier transform
of the temperature anisotropies in the small field limit. They
quantify generic non-Gaussian structure and complement the power
spectrum in characterizing the sampled distribution function of
a data set.
}
%
Recent years have seen scattered attempts at quantifying
non-Gaussianity in cosmological data sets. It has become clear that
there are two main lines of attack. One can focus on speculated
sources of non-Gaussianity and design statistics which can best
discriminate between them and Gaussian counterparts. This has
been the approach in, for example, the study of cosmic defects
such as strings or texture. The other, less prejudiced,
approach is to define a general framework with which one
can quantify deviations from Gaussianity. The main example
of this strategy has been the $n-$point formalism which has
been applied in a variety of settings \cite{p}. The redundancy and inefficiency
of such a method makes it pressing to look for viable alternatives.

This second approach is in general too ambitious. One must
 define restrictions to make any form of quantitative analysis
tractable. One can do this by establishing requirements. 
We shall describe a formalism \cite{fm} which
\begin{itemize} 
\item preserves information, i.e. given $N$ independent 
pixels this will supply $N$ quantities, one of which is the power spectrum
\item is defined in Fourier space, and therefore tailored for
high resolution, small field, interferometric measurements
\end{itemize} 

In this setting it makes sense to
consider the Fourier transform: 
\begin{equation}\label{fourier}
  \frac{\Delta T({\bf x})}{T}
={\int {d{\bf k}\over 2\pi}a({\bf k})e^{i{\bf k}\cdot{\bf x}}}
\end{equation}
A Gaussian probability distribution function of the complex $a({\bf k}_i)$
in a ring of fixed $|k|$ is given by: 
\begin{eqnarray}
  F[a({\bf k}_i)]=
      {1\over (2\pi\sigma^2)^{m_k}}
        \exp{\left(-
          {1\over 2\sigma_k^2}\sum_{i=1}^{m_k}|a({\bf k}_i)|^2\right)}
\end{eqnarray}
where we have $2m_k=f_{sky}(2k+1)$ independent modes ($f_{sky}$ is the
fraction of the sky covered). 
Defining $a({\bf k}_i)=\rho_ie^{{\rm i}\phi_i}$ we
can work in terms of $m_k$ moduli $\rho_i$ and $m_k$ phases $\phi_i$.
The  $\{\rho_i\}$ may  be seen as 
Cartesian coordinates which we transform into polar coordinates.
These consist of a radius $r$ plus $m_k-1$ angles $\tilde\theta_i$
given by
\begin{equation}\label{pol}
  \rho_i=r\cos{\tilde\theta_i}\prod_{j=0}^{i-1}\sin{\tilde\theta_j}
\end{equation}
with $\sin{\tilde\theta_0}=\cos{\tilde\theta_{m_k}}=1$.
In terms of these variables the radius is related to the
angular power spectrum by $C(k)=r^2/(2m_k)$. In general the first 
$m_k-2$ angles $\tilde\theta_i$ vary between $0$ and $\pi$ and the 
last angle varies between 0 and $2\pi$.
However because all $\rho_i$ are positive all angles are in $(0,\pi/2)$.
In order to define $\tilde\theta_i$ variables 
which are uniformly distributed in 
Gaussian theories one may finally perform the transformation on each
$\tilde\theta_i$:
\begin{equation}
  {\theta_i}=\sin^{N_k-2i}(\tilde\theta_i)
\end{equation}
so that for Gaussian theories one has:
\begin{equation}
  F(r,{\theta}_i,\phi_i)={r^{N_k-1}e^{-r^2/(2\sigma_k^2)}
    \over 2^{m_k-1}(m_k-1)!}\times 1 \times\prod_{i=1}^{m_k}{1\over
    2\pi}
\end{equation}
The factorization chosen shows that all new variables are independent
random variables for Gaussian theories. $r$ has a $\chi^2_{N_k}$
distribution,
the ``shape'' variables $\theta_i$ are uniformly distributed
in $(0,1)$, and the phases $\phi_i$ are uniformly distributed in $(0,2\pi)$.

The  variables $\theta_i$ define a non-Gaussian shape spectrum,
the {\it ring spectrum}. 
They may be computed from ring moduli $\rho_i$  simply by
\begin{equation}
  {\theta}_i={\left(\rho_{i+1}^2+\cdots +\rho_{m_k}^2
      \over \rho_i^2\cdots +\rho_{m_k}^2\right)}^{m_k-i}
\end{equation}
They describe how shapeful the perturbations are. 
If the perturbations are stringy then
the maximal moduli will be much larger than the minimal moduli.
If the perturbations are circular, then all moduli will be roughly
the same. This favours some combinations of angles, which are
otherwise uniformly distributed. In general any shapeful picture
defines a line on the ring spectrum $\theta_i$.
A non-Gaussian theory ought to define a set of probable smooth
ring spectra peaking along a ridge of typical shapes.

We can now construct an invariant for each adjacent pair of
rings, solely out of the moduli. If we order the $\rho_i$ for each
ring, we can identify the maximum moduli. Each of these moduli
will have a specific direction in Fourier space; let 
 ${\bf k}_{max}$
and ${\bf k}^{'}_{max}$ be the directions where the maximal moduli
 are achieved.
The angle
\begin{equation}
  \psi(k,k')={1\over \pi}{\rm ang}({\bf k}_{max},{\bf k}^{'}_{max})
\end{equation}
will then produce an inter-ring correlator for the moduli, the
{\it inter-ring spectra}. This 
is uniformly distributed in Gaussian theories in $(-1,1)$. It gives
us information on how connected the distribution of power is between
the different scales. 

We have therefore defined a transformation from the original modes
into a set of variables $\{r,\theta,\phi,\psi\}$. The non-Gaussian
spectra thus defined have a  particularly simple distribution
for Gaussian theories.
We shall call perturbations for which the phases are not uniformly
distributed localized perturbations. This is because if perturbations
are made up of lumps statistically distributed but with well defined 
positions then the phases will appear highly correlated. We shall
call perturbations for which the ring spectra are not
uniformly distributed shapeful perturbations. 
This distinction is interesting as it is in principle
possible for fluctuations to be localized but shapeless, or more 
surprisingly, to be shapeful but not localized. Finally we shall call 
perturbations for which the inter-ring spectra are not uniformly
distributed, connected perturbations. This turns out to be one of
the key features of perturbations induced by cosmic strings. 
These three definitions
allow us to consider structure in various layers. White noise
is the most structureless type of perturbation. Gaussian fluctuations
allow for modulation, that is a non trivial power spectrum $C(k)$,
but their structure stops there.
Shape, localization, and connectedness constitute the three next
levels of structure one might add on. Standard visual structure
is contained within these definitions, but they allow for more
abstract levels of structure. 
%
%
%
%
%
%
%
%
%

\end{document}